\documentclass[aps,twocolumn,showpacs,preprintnumbers,amsmath,amssymb]{revtex4}
\usepackage{graphicx}
\usepackage{bm}
\usepackage{dcolumn}

\def\nat#1#2#3{Nature {\bf #1}, #2 (#3)}
\def\natp#1#2#3{Nat. Phys. {\bf #1}, #2 (#3)}

\def\rmp#1#2#3{Rev. Mod. Phys. {\bf #1}, #2 (#3)}
\def\prl#1#2#3{Phys. Rev. Lett. {\bf #1}, #2 (#3)}
\def\pra#1#2#3{Phys. Rev. A {\bf #1}, #2 (#3)}

\def\epjd#1#2#3{Eur. Phys. J. D {\bf #1}, #2 (#3)}

\def\pla#1#2#3{Phys. Lett. A {\bf #1}, #2 (#3)}

\def\jpb#1#2#3{J. Phys. B: At. Mol. Opt. Phys. {\bf #1}, #2 (#3)}

\def\noi{\noindent}
\def\bc{\begin{center}}
\def\ec{\end{center}}
\newcommand{\bea}{\begin{equation}}
\newcommand{\eea}{\end{equation}\noi}
\newcommand{\ber}{\begin{eqnarray}}
\newcommand{\eer}{\end{eqnarray}\noi}
\begin{document}
\title{Collapse of a Bose gas: kinetic approach}
\author{Shyamal Biswas}\email{sbiswas.phys.cu@gmail.com}
\affiliation{Department of Physics, University of Calcutta, 92 APC Road, Kolkata-700009, India}
\date{\today}
\begin{abstract}
We have analytically explored temperature dependence of critical number of particles for the collapse of a harmonically trapped attractively interacting Bose gas below the condensation point by introducing a kinetic approach within the Hartree-Fock approximation. The temperature dependence obtained by this easy approach is consisted with that obtained from the scaling theory.
\end{abstract}
\maketitle
\section*{1. Introduction}
For an ultracold Bose gas, inter-particle interaction is characterized by s-wave scattering length ($a_s$) which can be tuned arbitrarily by the Feshbach resonance method \cite{feshbach}. For attractive ($a_s<0$) interaction, a harmonically trapped Bose gas tends to increase its density in the central region of a trap. This tendency, of course, is opposed by the quantum and thermal fluctuations. If the number of atoms is greater than a critical number ($N_c$) then the central density increases strongly, and the zero-point and thermal fluctuations are no longer able to avoid the collapse of the gas. Consequently, the gas becomes unstable even for two body interaction.

The stability and collapse of the Bose-Einstein condensates with negative scattering lengths have already been observed in the clouds of ultracold $^7$Li \cite{rice} and $^{85}$Rb \cite{jila} for temperatures ($T$) close to zero or well below the condensation point ($T_c$). Soon after the observation, a number of theory for the collapse have been proposed for $T\rightarrow0$ \cite{baym,pitaevskii,kagan,ueda,adhikari,savage,yukalov}, and for $T>0$ \cite{houbiers,davis,mueller,biswas} as well. The remarkable one among these (theories) was given by Baym and Pethick \cite{baym}. They proposed a scaling theory for $T=0$, and one of us generalized their theory for $0\le T\le T_c$ within the Hartree-Fock (H-F) approximation \cite{biswas}. In the generalized theory, different parts of the free energy (grand potential) of our system were scaled by a parameter which reduces the length scale of the system as a result of attractive interaction; and a critical number for the collapse was eventually calculated from a critical condition of existence of a metastable minimum of the grand potential \cite{biswas}. In this brief report we will also calculate the same, but in a kinetic approach which is supposed to be the easiest way.

This time, for calculating the critical number, we will not start from the free energy, but will adopt a mere kinetic theory like approach based on the energy and pressure of the system. We will start from the H-F energy of the system, and will pick up the kinetic energy and interaction energy parts of the H-F energy. While the kinetic energy of the particles causes an outward pressure the attractive interaction causes an inward pressure. For critical number of particles, magnitudes of the two pressures would be the same. Beyond the critical number of particles, the inward pressure would be larger than the outward one, and as a consequence, the whole system would collapse. Thus we will calculate the critical number, and will show its temperature dependence. Our present technique is easier than that of the already existing theory \cite{biswas,baym} as because outward and inward pressures appear as the first order derivative of the two parts of energy with respect to the effective volume of the system, and the already existing technique involves a second order derivative of the grand potential with respect to the scaling parameter for obtaining the critical condition of existence of its metastable minimum.

\section*{2. Qualitative result}
Before going into the details of the kinetic approach, let us estimate the critical number by qualitative manner. Our system consists of a large number ($N$) of Bose particles each of which is a 3-D isotropic harmonic oscillator with angular frequency $\omega$ and mass $m$. The system, of course, is in thermodynamic equilibrium with its surroundings at temperature $T$. For $T\rightarrow0$, all the particles occupy the ground state, and the system can be well described by the ground state wave function $\Psi_0({\bf{r}})=\sqrt{\frac{N}{l^3\pi^{3/2}}}e^{-\frac{r^2}{2l^2}}$ in the position ({\bf r}) space, where $l=\sqrt{\hbar/m\omega}$ is the confining length scale of the oscillators \cite{rmp}. Thus for $T=0$, the density of the condensed particles is given by \cite{rmp}
\bea
n_0({\bf{r}})=\mid\Psi_0({\bf{r}})\mid^2=\frac{N}{l^3\pi^{3/2}}e^{-\frac{r^2}{l^2}}.
\eea
On the other hand, the number density of the excited particles (in absence of interaction) is given by \cite{pitaevskii-book,biswas}
\bea
n_T({\bf{r}})=\frac{1}{\lambda_T^3}g_{\frac{3}{2}}(e^{-\frac{m\omega^2r^2}{2k_BT}}),
\eea
where $\lambda_T=\sqrt{\frac{2\pi\hbar^2}{mk_BT}}$ is the thermal de Broglie wavelength, and $g_{\frac{3}{2}}(x)=x+x^2/2^{3/2}+x^3/3^{3/2}+...$ is a Bose-Einstein function of a real variable $x$.

Let us consider the attractive interaction potential as $V_{int}({\bf{r}})=g\delta^3({\bf{r}})$, where $g=-\frac{4\pi\hbar^2a}{m}$ is the coupling constant and $a=-a_s$ is the absolute value of the s-wave scattering length  \cite{pitaevskii-book,biswas1,biswas2}. Typical two body interaction energy for $N$ number of particles is $\sim N^2 g/2l^3$. For this interaction, the gas tends to increase the density at the central region of the trap. Well below $T_c$ (i.e. for $T\rightarrow 0$), this tendency is resisted by the zero-point motion of the atoms. At the critical situation, the energy ($3N\hbar\omega/2$) for the zero point motion must be comparable to the typical interaction energy. So, for $T\rightarrow0$, we must have $3N_c\hbar\omega/2\sim N_c^2 g/2l^3$ at the critical situation. From this relation we may have $\frac{N_ca}{l}\sim 1$.

On the other hand, for $0<T<T_c$, the typical energy of the system is $3\hbar\omega[kT/\hbar\omega]^4\zeta(4)\sim N^{4/3}\hbar\omega$ \cite{rmp}. So, for this range of temperatures, we must have $N_c^{4/3}\hbar\omega\sim N_c^2 g/2l^3$ at the critical situation. From this relation, we may have $\frac{N_ca}{l}\sim[\frac{l}{a}]^{1/2}>1$.

However, near $T_c$, the length scale of the system is $L_{T_c}\sim l\sqrt{\frac{k_BT_c}{\hbar\omega}}\sim lN^{1/6}$ \cite{rmp} so that we can write $N_c^{4/3}\hbar\omega\sim N_c^2 g/2L_{T_c}^3$ at the critical situation, and consequently, we may have $\frac{N_ca}{l}\sim [\frac{l}{a}]^5\gg 1$.

\section*{3. Quantitative result from scaling theory}
From the above qualitative arguments we have got $\frac{N_ca}{l}\propto 1$ for $T\rightarrow0$, $\frac{N_ca}{l}\propto [\frac{l}{a}]^{1/2}$ for $0<T<T_c$, and $\frac{N_ca}{l}\propto [\frac{l}{a}]^{5}$ for $T=T_c$. The proportionality constants were determined by a scaling theory within H-F approximation \cite{biswas}. The scaling theory gives the proportionality constants as 0.671 for $T=0$ \cite{baym,biswas}, $1.210 \frac{(T/T_c)^6}{(1-(T/T_c)^3)^3}$ for $0<T<T_c$ \cite{biswas}, and 2.253 for $T=T_c$ \cite{biswas}.

In the following, we will also calculate the same but in a different (kinetic) approach. This approach is supposed to be the easiest one. And, we want to know whether the easiest approach reproduces the similar results.

\section*{4. Kinetic approach}
Within the H-F approximation we have the expression of energy functional as \cite{pitaevskii-book,biswas}
\begin{eqnarray}
E&=&\int d^3{\bf{r}}\bigg[\frac{\hbar^2}{2m}n_0\mid \nabla\phi_0\mid^2+\sum_{i\neq 0}\frac{\hbar^2}{2m}n_i\mid \nabla\phi_i\mid^2\nonumber\\&&+V(r)n_0({\bf{r}})+V({\bf{r}})n_T({\bf{r}})+\frac{g}{2}n_0^2({\bf{r}})\nonumber\\&&+2gn_0({\bf{r}})n_T({\bf{r}})+gn_T^2({\bf{r}})\bigg],
\end{eqnarray}
where $\{\phi_i\}$ represents the set of normalized H-F wavefunctions and $\{n_i\}$ represents their occupations. To evaluate the above energy functional of eqn.(3) we have to know $n_0({\bf{r}})$ and $n_T({\bf{r}})$ for the interacting case. But, for the simplicity of the calculation, we keep the forms of $n_0({\bf{r}})$ and $n_T({\bf{r}})$ as in eqns.(1) and (2) unaltered \cite{biswas,pitaevskii-book}. Thus we evaluate the energy functional in eqn.(3) as \cite{comments}
\begin{eqnarray}
E(t)=[(1+1)c_1(t)-c_2(t)]N\frac{\hbar^2}{ml^2},
\end{eqnarray}
where $t=T/T_c$,
\begin{eqnarray}
c_1(t)=\frac{3}{4}(1-t^3)+\frac{3}{2}\frac{N^{1/3}\zeta(4)t^4}{[\zeta(3)]^{4/3}},
\end{eqnarray}
\begin{eqnarray}
c_2(t)&=&\frac{1}{\sqrt{2\pi}}\frac{Na}{l}[1-t^3]^2+\sqrt{\frac{8\zeta(3/2)}{\pi}}\frac{N^{1/2}a}{l}t^{3/2}[1-t^3]\nonumber\\&&+S'\sqrt{\frac{2}{\pi[\zeta(3)]^3}}\frac{N^{1/2}a}{l}t^{9/2},
\end{eqnarray}
and $S'=\sum_{i,j=1}^{\infty}\frac{1}{(ij)^{3/2}(i+j)^{3/2}}\approx 0.6534$.
From eqn.(4) we get the kinetic energy of the system as
\bea
E_k(t)=c_1(t)N\frac{\hbar^2}{ml^2},
\eea
and the interaction energy of the system as
\bea
E_{int}(t)=-c_2(t)N\frac{\hbar^2}{ml^2}=-c_3(t)N^2a\frac{\hbar^2}{ml^3}
\eea
where \begin{eqnarray}
c_3(t)=c_2(t)\frac{l}{Na}.
\end{eqnarray}

For $0 \le T<T_c$, the effective volume of the system is given by $V=4\pi l^3/3$. The outward pressure is $P_{out}=\frac{\partial E_k}{\partial V}$, and the inward pressure is $P_{in}=\frac{\partial E_{int}}{\partial V}$. For the critical number of particles, $P_{in}$ would be equal and opposite to $P_{out}$. From this equality relation we can write the expression of the critical number as
\bea
N_c=\frac{2c_1(t)}{3ac_3(t)}\bigg(\frac{3V}{4\pi}\bigg)^{1/3}.
\eea

\subsection{Result for $T\rightarrow0$}
Eqn.(10) gives the expression of the critical number as
\bea
\frac{N_ca}{l}=\frac{2c_1(0)}{3c_3(0)}=\sqrt{\frac{\pi}{2}}=1.253 \ \ \text{for} \ \ T\rightarrow0.
\eea
Now we see that our new result for $T\rightarrow0$, is approximately two times larger than the analytic result ($0.671$) obtained from the scaling theory \cite{baym,biswas}. The numerical simulation and experimental results, for $T\rightarrow0$, however, are $0.575$ \cite{ruprecht} and $0.459$ \cite{jila} respectively.

\begin{figure}
\includegraphics{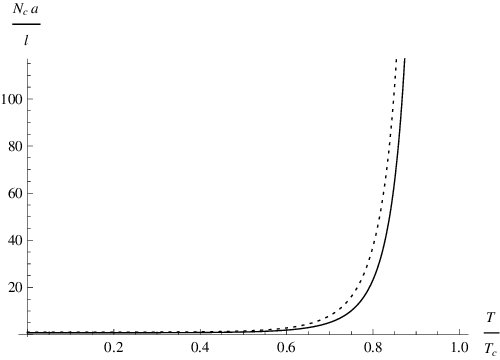}
\caption {Solid line follows form eqn.(12) and represents the plot for critical number of particles (in units of $\frac{l}{a}$) with respect to temperature (in units of $T_c$). Here $\frac{l}{a}= 0.0066$ \cite{jila}. Dotted line represents the results of the scaling theory \cite{biswas}.}
\end{figure}

\subsection{Results for $0<T<T_c$}
To achieve the Bose-Einstein condensation, the necessary condition is such that $a/l\ll 1$ \cite{pitaevskii-book}. Thus for $0<T\le T_c$, we can write $c_1(t)\approx\frac{3}{2}\frac{N^{1/3}\zeta(4)t^4}{[\zeta(3)]^{4/3}}$ from eqn.(5). Similarly, for $0<T\le T_c$, we can write $c_3(t)\approx\frac{1}{\sqrt{2\pi}}[1-t^3]^2$ from eqns.(9) and (6). Consequently, for $0<T<T_c$, we get the expression of $N_c$ from eqn.(10) as $\frac{N_ca}{l}\approx\frac{2\times \frac{3}{2}\frac{N_c^{1/3}\zeta(4)t^4}{[\zeta(3)]^{4/3}}}{3\frac{1}{\sqrt{2\pi}}[1-t^3]^2}$. From this relation we can write
\bea
\frac{N_ca}{l}=0.779\bigg[\frac{l}{a}\bigg]^{1/2}\frac{t^6}{(1-t^3)^3}+\frac{0.750}{1-t^3}+ {\it{O}}\bigg(\frac{a}{l}\bigg)^{1/4}
\eea
for $0<T<T_c$ up to a few leading orders in $a/l$.
We plot the right hand side of eqn.(12) with respect to $T$ in FIG. 1 for $a/l=0.0066$ \cite{jila}. Now, we see that the nature of the temperature dependence of the critical number is the same as that obtained from the scaling theory \cite{biswas}.

\subsection{Result for $T=T_c$}
For $T=T_c$, the length scale of the system is $L_{T_c}=lN^{1/6}$ \cite{pitaevskii-book,biswas}, and the effective volume of the system is given by $V=4\pi l^3 N^{1/2}/3$. For this reason, the critical condition would be the same as that in eqn.(10) except the factor $V$ which is to be replaced by $\frac{V}{N_c^{1/2}}$. So, for $T=T_c$, the critical number would be obtained from
\bea
N_c=\frac{2c_1(1)}{3ac_3(1)}\bigg(\frac{3V}{4\pi N_c^{1/2}}\bigg)^{1/3}.
\eea
Putting the values of $c_1(1)$ and $c_3(1)$ into the eqn.(13) we get
\bea
\frac{N_ca}{l}=96.062\bigg[\frac{l}{a}\bigg]^5 \ \ \text{for} \ \ T=T_c.
\eea
Now we see that the critical number increases dramatically as the condensation point is approached.

Difference between our result and the scaling result \cite{biswas,baym} is appreciable as because the scaling theory deals with all the terms of H-F energy, and the scaled kinetic energy although is greater than the scaled potential energy yet comparable in particular for $T\rightarrow0$. On the other hand, potential energy of the harmonic trap has not been considered in the analysis of our kinetic approach.

\section*{5. Conclusion}
It is to be mentioned that we have considered the collapse of the condensate as well as the thermal cloud of the Bose gas only for short ranged attractive interaction. Collapse of condensates for short ranged ranged repulsive and long ranged attractive interactions have also been investigated experimentally \cite{koch} and theoretically \cite{ticknor,haldar} as well.

Initially we explained the physics for the collapse of the attractive atomic Bose gas. Then we qualitatively estimated the critical number of particles for the collapse for $0\le T\le T_c$. Finally we calculate the same by a kinetic approach within the Hartree-Fock approximation. Our calculations support the qualitative estimations and the quantitative results obtained from a previous scaling theory \cite{biswas}.

Although the scaling theory was a more rigorous one yet the nature of the temperature dependence of the critical number obtained by our kinetic theory (in FIG. 1) is similar to that obtained from the scaling theory. These two theories are equal a priory and self consistent with respect to the basic physics of the collapse.

The scaling result, for $T\rightarrow0$, is closer to the experimental data \cite{jila} which, however, is not available for the entire regime $0<T\le T_c$. The power of our approach is the simplicity. Future experiment for $0<T\le T_c$ may justify our approach.

\section*{Acknowledgment}
This work has been sponsored by the University Grants Commission [UGC] under D.S. Kothari Postdoctoral Fellowship Scheme [No.F.4-2/2006(BSR)13-280/2008(BSR)]. Useful discussions with J.K. Bhattacharjee of SNBNCBS are gratefully acknowledged. Financial support and hospitalities of IACS are also acknowledged for an initial part of this work.


\begin{thebibliography}{99}
\bibitem{comments}It can be obtained from \cite{biswas} by putting $\nu=1$, where $\nu$ is the scaling parameter.
\end{thebibliography}
\end{document}